# New stable, explicit, first order method to solve the heat conduction equation

Endre Kovács

University of Miskolc, Department of Physics,

fizendre@uni-miskolc.hu

Highlights

- New first order scheme to solve heat conduction problems.
- Unconditionally stable method for the heat equation, thus no stepsize requirements.
- Explicit and one-step, easy to implement and parallelize.
- Handy to apply for any number of space dimensions and grid type.

**Abstract:** We introduce a novel explicit and stable numerical algorithm to solve the spatially discretized heat or diffusion equation. We compare the performance of the new method with analytical and numerical solutions. We show that the method is first order in time and can give approximate results for extremely large systems faster than the commonly used explicit or implicit methods.

## 1. Introduction and the studied problem

It is well known that the simplest Fourier-type heat conduction phenomena are described by the heat equation, which is a second-order parabolic partial differential equation (PDE), with the following form:

$$\frac{\partial T}{\partial t} = \alpha \Delta T + q,$$

where $\alpha = k/(c\rho)$ is the thermal diffusivity, $q$, $k$, $c$, and $\rho$ is the volumetric intensity of heat sources (radiation, chemical reactions radioactive decay, etc.), heat conductivity, specific heat and (mass) density, respectively.

The PDEs for real-life problems can rarely be solved analytically. The procedure of numerical solution begins with the discretization of the space variables. One divides the whole spatial domain into smaller cells, during which (in case of the heat equation) one has to calculate three quantities for each cell. The first one is the heat capacity of the cell: $C = c \cdot m = c\rho V$ [J/K], where $m$ is the mass, $V$ is the volume of the cell. Now the (thermal) energy of a cell can be expressed as $C_i \cdot T_i$, where $T_i$ is the average temperature of the cell. The second quantity is the heat/thermal conductance $U$, which can be approximated as

$$U_{ij} \approx k \frac{A_{ij}}{d_{ij}} \left[\frac{W}{K}\right],$$

where $A_{ij}$ is the surface area between the two cells $i$ and $j$, while $d_{ij}$ is the distance between the centres of the cells. The third quantity is $Q_i$, the heat source term:

$$Q_i = \frac{1}{C_i} \int_{V_i} q\, dV$$

After spatial discretization according to the usual second order [central difference formula] for the second derivatives, we obtain an [ODE] system which gives the time derivative of each temperature:

$$\frac{dT_i}{dt} = \sum_{j=neigh} \frac{U_{ij}}{C_i}(T_j - T_i) + Q_i,$$

where the summation is going over the neighbours of the cell. To help the reader to visualize, we present the arrangement of the variables in Fig 1. for a 2D system of 3 cells.

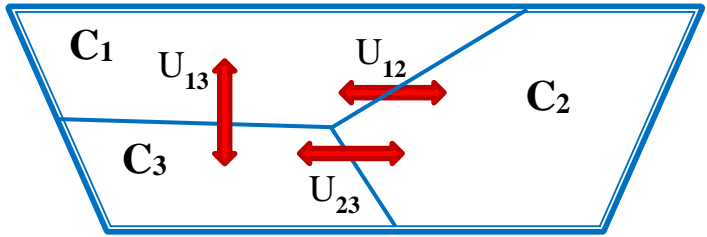

Figure 1 Notations in the case of three cells. The outer double line represents thermal isolation. We stress that the shape and arrangement of the cells are not necessarily regular.

The ODE system in a matrix form for this small system:

$$\frac{d\vec{T}}{dt} = \begin{pmatrix} -\dfrac{U_{12}}{C_1} - \dfrac{U_{13}}{C_1} & \dfrac{U_{12}}{C_1} & \dfrac{U_{13}}{C_1} \\ \dfrac{U_{12}}{C_2} & -\dfrac{U_{12}}{C_2} - \dfrac{U_{23}}{C_2} & \dfrac{U_{23}}{C_2} \\ \dfrac{U_{13}}{C_3} & \dfrac{U_{23}}{C_3} & -\dfrac{U_{13}}{C_3} - \dfrac{U_{23}}{C_3} \end{pmatrix} \vec{T} + \begin{pmatrix} Q_1 \\ Q_2 \\ Q_3 \end{pmatrix}$$

Suppose now that one quickly needs approximate results about the temperature distribution as a function of time in an extremely large and complicated system, where the physical properties of the material like the specific heat and the thermal diffusivity widely vary from point to point. In this case the size of the matrix is huge, while the magnitude of the matrix elements and therefore the eigenvalues have a range of several orders of magnitude, which means it is a severely stiff system. Which method can be recommended to solve this problem?

Conventional explicit methods are inappropriate because of unacceptably small timesteps due to stiffness. On the other hand, implicit methods require the solution of an algebraic equation system at each time-step, which is slow because of the necessity to store and handle huge matrices. Moreover, parallelization of implicit methods are nontrivial. The explicit methods with better stability properties like Runge–Kutta–Chebyshev, ADE (Alternating Direction Explicit), Hopscotch or Dufort-Frankel methods usually have other disadvantages: they can hardly be applied for irregular grids, they can be complicated to code, or they can only be conditionally consistent [1-4].

We have to conclude that conventional methods provide no convenient solution. In order to solve these systems more effectively, we started to elaborate a family of fundamentally new explicit methods. The simplest version for transient problems (i.e. when $q \equiv 0$) was already published [5]

## 2. The proposed method

Now we introduce the core method to solve the ODE system

$$\frac{d\vec{T}}{dt} = M\vec{T} + \vec{Q} \qquad (1)$$

through the following two steps:

1. We make a simplification: when we calculate the new value of a variable $T_i$, we neglect that other variables are also changing during the timestep. It means that we consider $T_j$ a constant if $j \neq i$, thus we can call it "**constant-neighbour** method". So we have to solve uncoupled, linear ODEs:

$$T_i' = a_i + m_{ii}T_i \qquad (2)$$

where $a_i = \sum_{j \neq i} m_{ij}T_{0,j} + Q_i$ and $m_{ii} = -1/\tau_i$ are considered as constants. We introduced $\tau_i = C_i / \sum_{j=neigh} U_{ij}$, which is the characteristic time of the cell, while $T_{0,j}$ are the initial temperatures.

2. We solve the obtained equations analytically. The analytical solution of eq. (2) at the end of the timestep is the following:

$$T_i(t+h) = T_i(t) \cdot e^{-\frac{h}{\tau_i}} + a_i \tau_i \cdot \left(1 - e^{-\frac{h}{\tau_i}}\right) =$$

$$= T_i(t) \cdot e^{-\frac{h}{\tau_i}} + \left(\sum_{j \neq i} \frac{U_{ij}}{C_i} T_{0,j} \frac{C_i}{\sum_{j=neigh} U_{ij}} + Q_i \tau_i\right)\left(1 - e^{-\frac{h}{\tau_i}}\right)$$

Thus we suggest the following simple formula to obtain the values of $T$ at the end of the timestep using the values of $T$ only at the beginning of the timestep:

$$T_i(t+h) = \left\{T_i(t) \cdot e^{-\frac{h}{\tau_i}} + \frac{\sum_{j=neigh} U_{ij} T_{0,j}}{\sum_{j=neigh} U_{ij}}\left(1 - e^{-\frac{h}{\tau_i}}\right)\right\} + Q_i \tau_i \left(1 - e^{-\frac{h}{\tau_i}}\right) \quad (3)$$

The first two terms on the right hand side, in curly brackets describe the transient process. As physically justifiable, the temperature of each cell exponentially tends to the temperature of its neighbours: the new value of the variable $T_i$ is the weighted average of the old value of $T_i$ and its neighbours $T_j$. It is easy to see that the coefficients of the temperatures are nonnegative and the sum of them is 1. That is why the result is always bounded: the method cannot be unstable.

This method has the following advantages:
1) It is obviously explicit, one can calculate the new values without solving a system of equations or even without using matrices. It also implies that the process is easily parallelizable and even vectorizable.
2) It is unconditionally stable for the heat conduction equation.
3) It can be easily applied regardless of the number of space dimension, grid irregularity and inhomogeneity of the heat conduction medium.

We performed numerical tests on several systems, but here we present only two different examples.

## 3. Verification: comparison with an exact result

If we have only two variables and one heat source, eq. (e1) has the following analytical solution:

$$T_1(t) = T_{0,1} \exp\left(\frac{-t}{\tau}\right) + T_a\left(1 - \exp\left(\frac{-t}{\tau}\right)\right) + St + S\tau \frac{C_2}{C_1}\left(1 - \exp\left(\frac{-t}{\tau}\right)\right)$$

$$T_2(t) = T_{0,2} \exp\left(\frac{-t}{\tau}\right) + T_a\left(1 - \exp\left(\frac{-t}{\tau}\right)\right) + St - S\tau\left(1 - \exp\left(\frac{-t}{\tau}\right)\right)$$

where $\tau = \dfrac{C_1 \cdot C_2}{U(C_1 + C_2)}$, $T_a = \dfrac{T_{0,1} C_1 + T_{0,2} C_2}{C_1 + C_2}$ and $S = \dfrac{Q_1}{C_1 + C_2}$ is the common characteristic time, the final temperature without external source (weighted average of the initial temperatures) and the common increment of the temperatures due to the $Q_1$ source in one time unit, respectively. We obtained that results produced by our method fit the exact values very well at $t_{\text{fin}}$. An example for the errors as a function of the stepsize is presented in Fig. 2. It is well known that a method is called to be $p$th order if the local error is $O(h^{p+1})$, or (equivalently for normal systems) if the global error is $O(h^p)$. From the figure one can see that the global error decreases with the first power of the stepsize, thus one can conclude that this method is first order. The mathematical proof of this statement is presented in the Appendix.

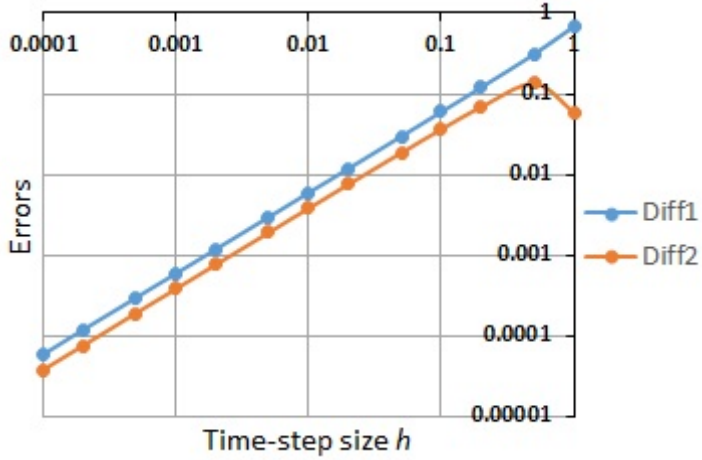

Figure 2 Absolute value of errors as the function of the timestep-size for
$U=1$, $C_1=5$, $C_2=1$, $T_{0,1}=10$, $T_{0,2}=0$, $Q_1=8$ and at final time $t_{FIN}=1$. Diff1 is for the first, Diff2 is for the second cell.

## 4. Comparison with numerical results for a stiff system

The second system is a rectangle-shaped lattice, $N_x$=100, $N_y$=50. A value $10^{(3-7 \cdot rand)}$ were given to the capacities $C_i$, and to the inverse conductances $1/U_{xi}$, $1/U_{yi}$ (the resistances), where *rand* is a random number generated by the MATLAB uniformly in the (0,1) interval for each quantity. It means that the capacities (the resistances) follow a log-uniform distribution between 0.0001 and 1000. The initial temperatures are uniformly zero, the sources have a uniform distribution between 0 and 100, $Q_i = 100 \cdot (1-rand)$. The task is to solve this system for the temperatures between $t_0$=0s and $t_{FIN}$=10s.

This is a seriously stiff problem, the stiffness ratio is $7.6 \cdot 10^{12}$. For the explicit Euler method (which is equivalent to the forward-time central-space FTCS scheme), the maximum possible timestep is $h_{MAX}^E = |2/\lambda_m| = 1.45 \cdot 10^{-8} s$, above this threshold instability necessarily occurs. Here $\lambda_m$ is (non-positive) eigenvalue of the matrix with the largest absolute-value.

For this system, not only the analytical solution is hopeless, but the widely used (conditionally stable) explicit methods are also inconvenient, as they would require at least a day to reach any result. Therefore, to provide us a reference solution, we used an implicit ode15s solver of MATLAB, which is variable-step, variable-order, based on the numerical differentiation formulas (NDFs) of orders 1 to 5, where the letter *s* indicates that the codes were designed especially for stiff systems. With strict error tolerance $(\text{'}RelTol\text{'}=10^{-8},\text{'}AbsTol\text{'}=10^{-7})$, our computer needs 677s to give a high precision solution using this routine.

With our "constant-neighbour" method, one timestep takes roughly 0.0004s. We found that if $h=0.0002$, then the produced results already fit quite well to the exact curve. The result is presented in Fig. 2. One can see that we managed to obtain a reasonable solution in $t_{FIN} \cdot 0.0004\text{s}/h \approx 20\text{s}$ (in the reality, 14s), much faster than the conventional explicit or implicit methods.

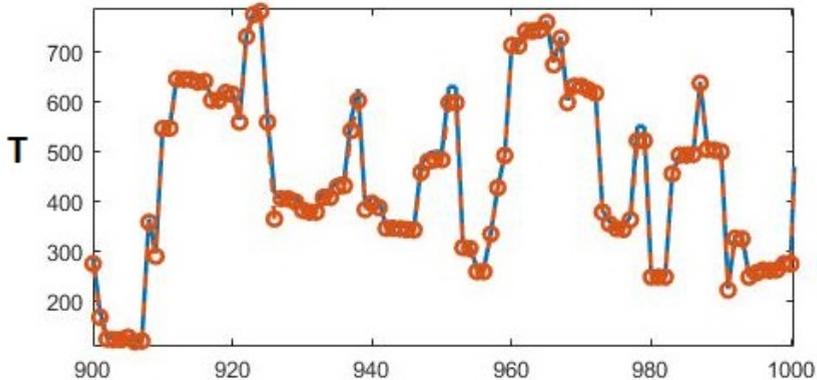

Figure 2 A randomly selected part of the graph of the temperature as a function of the space variable. The blue line is the high-precision solution while the orange circles are the values produced by our algorithm for $h=0.0002$.

In Table 1. we summarize some results obtained by MATLAB routines ode15s and ode23s and our method. We note that no matter how huge error tolerance is set to enhance speed, we were not able to obtain any results in 3 minutes by any MATLAB routines.

| method | Runtime | MaxD | SumD | SumEnD |
|---|---|---|---|---|
| ode15s | 181 | 7.7 | 4890 | 360684 |
| ode15s | 248 | 0.49 | 266.4 | 15828 |
| ode15s | 400 | 0.001 | 0.506 | 25.28 |
| ode23s | 2112 | 0.146 | 79.6 | 4725 |
| CN, $2 \cdot 10^{-4}$ | 14 | 340.9 | 38702 | 870038 |
| CN, $2 \cdot 10^{-5}$ | 142 | 36.65 | 3570 | 80143 |
| CN, $10^{-5}$ | 307 | 15.75 | 1715 | 38821 |
| CN, $5 \cdot 10^{-6}$ | 556 | 7.06 | 823 | 18783 |

Table 1. Performance of 3 different solvers, CN is for "Constant-Neighbour". The first one, MaxD is the maximum deviation (absolute value of the difference) from the reference solution. The second one, SumD is the sum of the deviations for all of the cells. The third one is the same but weighted with the capacities, $SumEnD = \sum_{i=1}^{N} C_i \cdot |T_i(t = t_{FIN}) - T_i(t = t_0)|$ to give us the error in term of energy.

From the data one can see that when larger error tolerance is set, our method is at least comparable with the standard solvers. On the other hand, if one needs a high-precision solution, then we could not recommend it, but first order methods are not for this purpose, anyway. However, we emphasize the following facts:

- This is still a small system with 5000 cells, and for larger number of cells implicit methods have more serious drawbacks. We note that the reason we have not used larger systems for testing purposes is that our computer cannot solve them by implicit methods because of the huge memory requirements.

- This is the simplest, "raw" version of our method, without any optimization, adaptive stepsize control, parallelization or vectorization, which could immensely enhance computing speed.

On Fig. 3. we present the three different kinds of error as a function of the stepsize $h$. One can see that for smaller stepsize, the errors are decreasing slightly faster than the stepsize, thus it underpins that the convergence-rate of the method is (at least) one. We stress that the infamous phenomenon of order reduction [6] has never been observed in case of our method. The right side of the diagram also reinforces the statement that the method is stable, as the error does increase quite slowly for increasing stepsize.

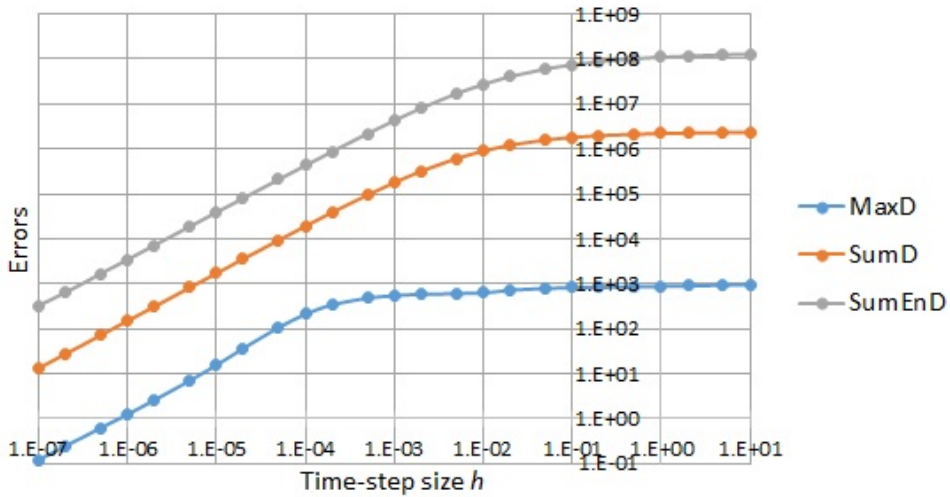

Figure 3 Different kind of errors as the function of the timestep-size. For the definition of the error-quantities, see the caption of Table 1.

## 5. Summary

We presented a new numerical algorithm to solve the heat conduction or diffusion equation with external sources. This method is explicit,

unconditionally stable and first order in time. We illustrated the performance of the method for a simple, analytically soluble case and in the case of a 2D system with highly inhomogeneous random parameters. The obtained data suggest that if quick results are required for a huge number of cells, our method has a remarkable advantage, even without parallelization. We have also proven analytically that the method is first order in time. We note that we are currently working on the second order version, which will be published elsewhere.

## 6. Acknowledgement

I would like to thank András Gilicz for introducing me to this way of thinking and for the long years of collaboration.

**8. Appendix: The proof that the method is first order in time**

Using the power series form of the exponential function, the exact solution of (1) is the following:

$$\vec{T}(t) = e^{Mt}\vec{T}_0 + \left(e^{Mt} - 1\right)M^{-1}\vec{Q} =$$
$$= \left(1 + Mt + M^2\frac{t^2}{2} + M^3\frac{t^3}{3!} + \ldots\right)\vec{T}_0 + \left(t + M\frac{t^2}{2} + M^2\frac{t^3}{3!} + \ldots\right)\vec{Q}.$$

The 0th and first order terms in the exact solution at $t=h$:

$$T_i(h) = T_{0,i}\left(1 + m_{ii}h\right) + h\sum_{j\neq i}m_{ij}T_{0,j} + Q_i h = T_{0,i}\left(1 - \frac{h}{\tau_i}\right) + h\sum_{j=neigh}\frac{U_{ij}}{C_i}T_{0,j} + Q_i h,$$

where we used the fact that $U_{ij} \neq 0$ iff the two cells are neighbours.

Let us compare it to our "Constant Neighbour" solution. From (3) we obtain

$$T_i(h) = T_{0,i}\left(1 - \frac{h}{\tau_i} + O(h^2)\right) + \left(\frac{\sum_{j=neigh}U_{ij}T_{0,j}}{\sum_{j=neigh}U_{ij}} + Q_i\tau_i\right)\left(1 - 1 + \frac{h}{\tau_i} - O(h^2)\right)$$

The 0th and 1st order terms:

$$T_i(h) \approx T_{0,i}\left(1 - \frac{h}{\tau_i}\right) + \left(\frac{\sum_{j=neigh}U_{ij}T_{0,j}}{\sum_{j=neigh}U_{ij}} + Q_i\tau_i\right)\frac{h}{\tau_i} =$$
$$= T_{0,i}\left(1 - \frac{h}{\tau_i}\right) + h\frac{\sum_{j=neigh}U_{ij}T_{0,j}}{\sum_{j=neigh}U_{ij}}\frac{\sum_{j=neigh}U_{ij}}{C_i} + Q_i h$$

After simplification we obtain that the difference between the exact solution and our result is second order in time.